%% file: main.tex
\documentclass[a4paper,11pt]{article}

\usepackage[margin=1.5in]{geometry}    
\usepackage[english]{babel}
\usepackage{tikz}
\usepackage{quantikz}
\usetikzlibrary{positioning}
\usepackage{amsmath, amssymb}
\usepackage{booktabs}
\usepackage{hyperref}
\usepackage{here}
\usepackage{fancyvrb}
\usepackage{fvextra}
\title{Molecular Algebraic Geometry: Electronic Structure of H$_3^+$ as Algebraic Variety}

\author{Ichio Kikuchi$^{1}$, Akihito Kikuchi$^{2}$\footnote{akihito\_kikuchi@gakushikai.jp (The corresponding author; a visiting researcher in IFQT)}  \\
        \small $^{1}$Internationales Forschungszentrum f\"ur Quantentechnik\\
        \small $^{2}$International Research Center for Quantum Technology, Tokyo \\
}

\date{\today}    

\begin{document}\maketitle
\begin{abstract}
In this article, we demonstrate the restricted Hartree-Fock electronic structure computation of the molecule $H_3^+$ through computational algebra. We approximate the Hartree-Fock total energy by a polynomial composed of LCAO coefficients and atomic distances so that the minimum is determined by a set of polynomial equations. We get the roots of this set of equations through the techniques of computational algebraic geometry, namely, the Gr\"obner basis and primary ideal decomposition. This treatment enables us to describe the electronic structures as algebraic varieties in terms of polynomials. 
\end{abstract}

\section{
Introduction
}
We have developed an algebraic method of quantum chemistry \cite{KIKUCHI2013}, whereby we conduct computations in the following way:
\begin{itemize}
\item The molecular integrals are given by analytic formulas. This is possible when we use analytic atomic orbitals such as Gaussian-Type or Slater-Type orbitals (GTO or STO). By Tailor expansion, the molecular integrals are approximated by polynomials.
\item The total energy is a polynomial composed of those molecular integrals and the undetermined coefficients of LCAO. The ortho-normalization conditions are similarly treated.
\item We compose the objective function from the total energy and the ortho-normalization condition with the Lagrange multipliers (which are the orbital energies)
\item By differentiation, we obtain the set of polynomial equations that gives the optima.

\item To solve it, we apply the method of computer algebra, where Gr\"obner bases and the primary ideal decomposition play central roles to get the quantum eigenstates. 

\item One might use the term {\it Molecular Algebraic Geometry} to refer to this algebraic computational scheme for molecular orbital theory.
\end{itemize}

Up to now, we have reported the results of the hydrogen molecule, using STO and n-GTO models ( \cite{KIKUCHI2013} and \cite{kikuchi2022ab}).   

In this article, we report the result of the H$_3^+$ molecule. 

The trihydrogen has been the object of study for decades
in various fields from its discovery, since this molecule is the simplest triatomic molecule and abundant in the universe; on account of its structural simplicity, this molecule is a suitable object of benchmark problem of quantum chemistry \cite{stevenson1937structure,hirschfelder1938energy,king1979theory,herbst2000astrochemistry,tashiro2002quantum,kokoouline2003theory,oka2005hot,foroutan2009chemical,pavanello2009calculate,jadhav2020theoretical,kannan2021rydberg}. The present study goes back to the basics of quantum chemistry, whereby it demonstrates how the computational algebraic
geometry shall serve as an alternative approach of quantitative ab-initio computation.  Because of the limitation of the current computational powers, we are obliged to work in literally restricted models (RHF). However, for the pedagogical purpose, this serves well, since it shall show the steps of ab-initio quantitative computation of the ground states, the excited states, and the virtual orbitals. The feature of the method is to describe the quantum states as a series
of algebraic varieties, obtained from computations. An additional example shall demonstrate that algebraic geometry is useful in judging the existence of solutions; this sort of judgment is impossible for the conventional method that utilizes solely numerical tools. We will elaborate on how to compute polyatomic
molecules through the computation of H$_3^+$.
We elaborate on how to treat polyatomic systems.

The preliminaries for this kind of algebraic computations are reviewed in \cite{kikuchiquantum2021, kikuchi2019computational} and the references cited therein. If you would like to comprehend the required mathematics rigorously, you should consult \cite{decker2006computing, SOTTILE,cox2006using,cox2013ideals, COX,ene2011gr}.
The basic analytic formulas concerning Hartree-Fock computations by GTO are given in \cite{szabo2012modern}. Moreover, at our GitHub \cite{OURGITHUB}, we show short programs used in this study. We compute the symbolic formulas by SymPy package of Python \cite{10.7717/peerj-cs.103}, and we make use of the computer algebra system SINGULAR \cite{SINGULAR} for the computations in the algebraic geometry.

\section{
Model description
}
We assume the equilateral triangle model of H$_3^+$ (with the bond length $R$) and put atomic bases at three centers A, B, and C. The trial wave function is given by
\[
\psi(r)=x\cdot\phi(r,A)+y\cdot\phi(r,B)+z\cdot\phi(r,C)
\]
with the orbital energy $e$. For brevity, we write the wavefunction by $(x,y,z)$.
We use the STO-3G basis set for the atomic bases:
\[
\phi(r)=\sum_{i=1}^3 d_i \exp (-b_i r^2).
\]
In that formula, the parameters are given as
\[
b_i= a_i \zeta^2, 
\]
and
\[
d_i= c_i \left( \frac{2b_i}{\pi} \right)^{3/4}
\]
for $i=1, 2, 3$. The numeral data are given in Tables \ref{STO3G} and \ref{STO3GZ}.

\begin{table}[ht]
\begin{center}
\begin{tabular}{lrr}
\toprule
i &    c(i) &    a(i) \\
\midrule
1 &  0.4446 &  0.1098 \\
2 &  0.5353 &  0.4058 \\
3 &  0.1543 &  2.2277 \\
\bottomrule
\end{tabular}
\caption{
The exponents and coefficients of STO-3G of H.
}
\label{STO3G}
\end{center}
\end{table}

\begin{table}[ht]
\begin{center}
\begin{tabular}{lrr}
\toprule
$\zeta$ &  1.24 \\
\bottomrule
\end{tabular}
\caption{
The scale factor of STO-3G of H.
}
\label{STO3GZ}
\end{center}
\end{table}

We prepare the following molecular integrals for every possible combination of atomic bases (indexed by P, Q, K, L, and so on), wherein the summation for those indices is taken over the centers of the orbitals A, B, and C.

Overlap integrals:
\[
S_{PQ}=\left(\phi(P)\middle| \phi(Q)\right)
\]

Kinetic integrals:
\[
K_{PQ}=\left(\phi(P)\middle|\ -\frac{1}{2}\nabla^2\ \middle| \phi(Q)\right)
\]

One-center integrals (namely, the nuclear potentials):
\[
V_{PQ,X}=-\left(\phi(P)\middle|\ \frac{1}{\mid r-X\mid }\ \middle| \phi(Q)\right)
\]

Two-center integrals:
\[
[PQ\mid XY]=\int dr_1\ dr_2\ \phi(r_1,P) \phi(r_1,Q)\ \frac{1}{| r_1-r_2| }\ \phi(r_2,X)\phi(r_2,Y)
\]
 
The skeleton part of the Fock matrix: 
\[
H_{PQ}=K_{PQ}+V_{PQ,A}+V_{PQ,B}+V_{PQ,C}
\]

The density matrix:
\[
D=2\begin{pmatrix}
    x^2 & xy &xz\\
    xy  & y^2 &yz\\
    xz  & xy & z^2
\end{pmatrix}
\]

The electron-electron interaction part of the Fock matrix:
\[
G_{PQ}=\sum_{K,L}D_{KL}[PQ\mid KL]-0.5D_{KL}[PL\mid QK]
\]

The Fock matrix:
\[
F_{PQ}=K_{PQ}+G_{PQ}+V_{PQ,A}+V_{PQ,B}+V_{PQ,C}
\]

The total energy (with the normalization condition and the nuclear-nuclear repulsion):

\[
E_{tot}=\frac{1}{2}\sum_{i,j} D_{ij}\cdot(H_{ij}+F_{ij}) -2 e \left(\sum_{i,j} \frac{1}{2} D_{ij}S_{ij}-1\right) +\frac{3}{R}
\]
 
The restricted Hartree-Fock computation by PySCF \cite{PYSCF} gives the reference data of the electronic structure of H$_3^+$  presented in Table. \ref{QCHEMH3P}
\begin{table}[H]
    \centering
    \begin{tabular}{l l}\hline
        Orbital energy of the occupied orbital ($E_{\rm h}$) &  -1.1991769643185\\
        Total energy ($E_{\rm h}$)   & -1.24233050684767\\\hline
    \end{tabular}
    \caption{
    The result of RHF computation for H$_3^+$ through PySCF. The computation was done at the interatomic distance of 0.9 \AA.
    }
    \label{QCHEMH3P}
\end{table}
\section{
Restricted Hartree-Fock computation
}
\label{SecRHF}

We set $R$ at 0.9 \AA (1.70 $a_0$) and fix the values of molecular integrals. Then the objective function is the polynomial of $(x,y,z,e)$. To avoid the numerical error in the symbolic computation conducted hereafter, we would like to work with integers. To this end, we truncate the digits at a finite length and represent them as rational numbers with the denominators of the powers of ten. By multiplying the objective functions with the common denominator of those rational numbers, we get the polynomial objective function with integer coefficients. 

In the case of H$_3^+$, the polynomial objective function
is given by
\begin{verbatim}
OBJ=-20000*e*x**2 - 22288*e*x*y - 22288*e*x*z
- 20000*e*y**2 - 22288*e*y*z - 20000*e*z**2
+ 20000*e + 7746*x**4 + 14316*x**3*y + 14316*x**3*z
+ 18392*x**2*y**2 + 26758*x**2*y*z + 18392*x**2*z**2 
- 31781*x**2 + 14316*x*y**3 + 26758*x*y**2*z 
+ 26758*x*y*z**2 - 44637*x*y + 14316*x*z**3
- 44637*x*z + 7746*y**4 + 14316*y**3*z 
+ 18392*y**2*z**2 - 31781*y**2 + 14316*y*z**3 
- 44637*y*z + 7746*z**4 - 31781*z**2 + 17639.
\end{verbatim}

From the minimum condition of the objective function, namely, through the partial differentiation by x,y,z, and e, we get a set of equations composed of the following polynomials:
\begin{verbatim}
I[1]=30984*x^3+42948*x^2*y+36784*x*y^2+14316*y^3
+42948*x^2*z+53516*x*y*z+26758*y^2*z+36784*x*z^2
+26758*y*z^2+14316*z^3-40000*x*e-22288*y*e
-22288*z*e-63562*x-44637*y-44637*z,

I[2]=14316*x^3+36784*x^2*y+42948*x*y^2+30984*y^3
+26758*x^2*z+53516*x*y*z+42948*y^2*z+26758*x*z^2
+36784*y*z^2+14316*z^3-22288*x*e-40000*y*e
-22288*z*e-44637*x-63562*y-44637*z,

I[3]=14316*x^3+26758*x^2*y+26758*x*y^2+14316*y^3
+36784*x^2*z+53516*x*y*z+36784*y^2*z+42948*x*z^2
+42948*y*z^2+30984*z^3-22288*x*e-22288*y*e
-40000*z*e-44637*x-44637*y-63562*z,

I[4]=-20000*x^2-22288*x*y-20000*y^2-22288*x*z
-22288*y*z-20000*z^2+20000.
\end{verbatim}

Those polynomials construct an ideal I. The Gr\"obner basis of the ideal I is composed of 9 polynomials (J[1]--J[9]) in the case of the lexicographic monomial ordering. Here we only show the concrete form of one of them, because the polynomials in the basis are too lengthy.
\begin{verbatim}
J[1] =36812798143709139749144994785898166628646912*e^5
+127499329072859314688739373550168339532546048*e^4
+168144765973871153972385957671411273274218496*e^3
+105507004438312720335072205511167338086667520*e^2
+31442209826685795119941751238920406402581000*e
+3570217123261998609632303137992216382496875.
\end{verbatim}

Instead of showing all polynomials, we give a brief description of the structures of the Gr\"obner basis in Table \ref{GBH3P}.
\begin{table}[htbp]
\begin{center}
    
\begin{tabular}{lll}
\toprule
{} &     variables &                                            formula \\
\midrule
1 &           \{e\} &  ()*e**5 + ()*e**4 + ()*e**3 + ()*e**2 + ()*e + () \\
2 &        \{e, z\} &  ()*e**4*z**2 - ()*e**4 + ()*e**3*z**2 - ()*e**... \\
3 &        \{e, z\} &  -()*e**4*z + ()*e**3*z**3 - ()*e**3*z + ()*e**... \\
4 &        \{e, z\} &  ()*e**4 - ()*e**3*z**2 + ()*e**3 - ()*e**2*z**... \\
5 &     \{y, z, e\} &  ()*e**4*y - ()*e**4*z + ()*e**3*y - ()*e**3*z ... \\
6 &     \{y, z, e\} &  -()*e**4*z - ()*e**3*y - ()*e**3*z - ()*e**2*y... \\
7 &     \{y, z, e\} &  -()*e**4 + ()*e**3*y*z + ()*e**3*z**2 - ()*e**... \\
8 &     \{y, z, e\} &  -()*e**4 + ()*e**3*y*z - ()*e**3 + ()*e**2*y*z... \\
9 &  \{y, z, e, x\} &  ()*e**4*z + ()*e**3*z + ()*e**2*z**3 + ()*e**2... \\
\bottomrule
\end{tabular}
\end{center}

\caption{
This table shows the nine polynomials in the computed Gr\"obner basis which determines the electronic states of H$_3^+$. As the integer coefficients are too lengthy, they are replaced with `()'. Each row shows the formula and the variables included in it.
}
\label{GBH3P}
\end{table}

The Gr\"obner basis has roots which is the same as that of the initially given set of polynomial equations.  The polynomials in the Gr\"obner basis are arranged in such a manner that the first of them includes only one variable e, and the polynomials which come later acquire other valuables e, z, y, and x in succession. This is the feature of the Gr\"obner basis computed with the lexicographic monomial ordering with $x < y < z <e$. The computation of Gr\"obner basis is an extension of Gaussian elimination in a matrix. The operation corresponding to the matrix row subtraction successively eliminates the monomials in the array of polynomials: the monomials located ahead in the ascending ordering shall be eliminated earlier, meanwhile, those located behind shall be retained. In the end, the monomials composed of the powers of e, which are the largest in the monomial ordering, give rise to a univariate polynomial in the Gr\"obner basis.  on account of the tidiness as a basis set, it is better to use the Gr\"obner basis for algebraic study.

We can advance more: the Gr\"obner basis is furthermore decomposed into the subsets which have a kind of triangular forms through the primary ideal decomposition. In the current study for H$_3^+$, a triangular subset is composed of four polynomials as follows: 
\begin{align*}
&t^{(i)}_1(e)\\
&t^{(i)}_2(e,z)\\
&t^{(i)}_3(e,z,y)\\
&t^{(i)}_4(e,z,y,z)\\
\end{align*}
The triangular subset is indexed by $i=1,..., M$, where $M$ is the size of the decomposed set. This triangulation shall be the most preferable form regarding numerical computation. The primary ideal decomposition for the quantum states of H$_3^+$ is given by
\begin{verbatim}
T[3]:
   _[1]=55883592e+67016387
   _[2]=7929z2-1250
   _[3]=y-z
   _[4]=x-z
T[1]:
   _[1]=19607184e+7819975
   _[2]=1107z3-1250z
   _[3]=1107y2+1107yz+1107z2-1250
   _[4]=x+y+z
T[2]:
   _[1]=33596906582185481529961979904e3
   +62671838378960619969574479872e2
   +37234589977346832313644870640e
   +6812527267627204196782565375
   _[2]=87306551258263729827909084908235z2
   -614312087531946676701270123361536e2
   -537304333596464593384375038355328e
   -139185734162956356442927812890825
   _[3]=331997850486797035543167841067099398038750y2
   -13015916170638763880890008092932038290847744yze2
   -13459096677815806942752983709330245786664480yze
   -2687870434415068556840506490446894349472000yz
   +11262181386982462508243034408486017150784000e2
   +12991002284039358505441186364021967659282000e
   +2951891604844851948796638575554228299190625
   _[4]=18444325027044279752398213392616633224375x
   +18444325027044279752398213392616633224375y
   -723106453924375771160556005162891016158208ze2
   -747727593211989274597387983851680321481360ze
   -149326135245281586491139249469271908304000z
T[4]:
   _[1]=33596906582185481529961979904e3
   +62671838378960619969574479872e2
   +37234589977346832313644870640e
   +6812527267627204196782565375
   _[2]=87306551258263729827909084908235z2
   +8514144339559070475415581022605696e2
   +9287598840997447194191617733989768e
   +2076320747503839544988065579439950
   _[3]=54922323119969587357485501941034906122670y
   +1008818016691184154099544389734581696829952ze2
   +1101539894576580693998431276321064052056416ze
   +286732811316955459193115321222624363426985z
   _[4]=54922323119969587357485501941034906122670x
   +1008818016691184154099544389734581696829952ze2
   +1101539894576580693998431276321064052056416ze
   +286732811316955459193115321222624363426985z
\end{verbatim}

We solve the equation through those formulas, by determining the unknown variables one by one. The result is given in Table \ref{H3QS}. Note that we omit the complex-valued solutions in this table.

\begin{table}[H]
\begin{center}
 \begin{tabular}{rrrrrrr}
\toprule
{}&x&y&z&e&Total Energy&Origin\\
\midrule
0&-0.3971&-0.3971&-0.3971&-1.1992&-1.2424&T[3]\\
1&0.3971&0.3971&0.3971&-1.1992&-1.2424&T[3]\\
2&0.0000&1.0626&-1.0626&-0.3988&0.2966&T[1]\\
3&1.0626&0.0000&-1.0626&-0.3988&0.2966&T[1]\\
4&0.0000&-1.0626&1.0626&-0.3988&0.2966&T[1]\\
5&-1.0626&0.0000&1.0626&-0.3988&0.2966&T[1]\\
6&1.0626&-1.0626&0.0000&-0.3988&0.2966&T[1]\\
7&-1.0626&1.0626&0.0000&-0.3988&0.2966&T[1]\\
8&1.2730&-0.5463&-0.5463&-0.3531&0.3254&T[2]\\
9&-0.5463&1.2730&-0.5463&-0.3531&0.3254&T[2]\\
10&0.5463&-1.2730&0.5463&-0.3531&0.3254&T[2]\\
11&-1.2730&0.5463&0.5463&-0.3531&0.3254&T[2]\\
12&0.5463&0.5463&-1.2730&-0.3531&0.3254&T[4]\\
13&-0.5463&-0.5463&1.2730&-0.3531&0.3254&T[4]\\
\bottomrule
\end{tabular}

\caption
{
The real solutions of the polynomial equations for H$_3^+$ are shown. Each row shows the values of $(x,y,z,e)$, the corresponding total energy $E$ ( the local optimum value of the objective function), and the triangular subset which yields the solution. 
}
\label{H3QS}
\end{center}
\end{table}


There are symmetries (or equivalences) concerning the entries of the solutions which are the consequence of the symmetry of the molecule. To understand their origin, let us assume that the molecule has the symmetry of $C_3$. Then we get the three types of LCAO wavefunctions which shall be the basis of the eigenfunctions as follows:
\[
x\cdot(1,1,1), x\cdot(1,\omega,\omega^2), x\cdot(1,\omega^2,\omega2) 
\]
with $\omega^3=1$. We require that the LCAO coefficients are real numbers for the usage of quantum chemistry. Then we choose the basis for the quantum states as follows.
\[
(x,x,x), (0,t,-t),(t,0,-t),(t,-t,0),(-2s,s,s),(s,-2s,s), (s,s,-2s)
\]  

In Table \ref{H3QS} and the triangular decomposition of the ideal, we observe the following items.
\begin{itemize}
\item 
The basis $(x,x,x)$ generates the solutions of No.0 and No.1. The triangular set T[3] encloses them, where the orbital energy is determined by a linear polynomial. 

\item 
The bases $(0,t,-t)$, $(t,0,-t)$, and $(t,-t,0)$ generate the solutions of No.2--7. The triangular set T[1] encloses them, where the orbital energy is determined by a linear polynomial.

\item 
The bases $(-2s,s,s)$, $(s,-2s,s)$, and $(s,s,-2s)$, mixed with the basis $(x,x,x)$, generate No.8--10, and No.11--13. Two triangular subsets T[2] and T[4] enclose those solutions. Each of these triangular subsets describes different electron configurations, although they have a common univariate polynomial that determines $e$.
\end{itemize}


\section{
The simultaneous optimization of the atomic coordinates and the electronic structure
}

The simulation of the Car-Parrinello type \cite{PhysRevLett.55.2471} is also one of the abilities of the present algebraic scheme.  In the Hartree-Fock computation in Section \ref{SecRHF}, we have fixed the interatomic distance at a certain numerical value. Instead, we could treat it as a variable so that we could determine the optimum. To this end, we approximate the analytic formula of the objective function by the Taylor expansion of $R$ at a fixed center $R_c =0.7$ \AA so that we could use the polynomial representation. We suppose that the molecule lies in the ground state, where the LCAO coefficients are given by $(x,y,z)=(x,x,x)$. Then we get the polynomial objective function:  
\begin{verbatim}
OBJ=-19*R**5*e*x**2 + 5162*R**5*x**4 - 2777*R**5*x**2 
- 1243*R**5 + 1619*R**4*e*x**2 - 56270*R**4*x**4 
+ 27032*R**4*x**2 + 12677*R**4 - 13645*R**3*e*x**2 
+ 231739*R**3*x**4 - 98114*R**3*x**2 - 53879*R**3 
+ 37479*R**2*e*x**2 - 381788*R**2*x**4 + 120507*R**2*x**2 
+ 122124*R**2 - 582*R*e*x**2 - 23533*R*x**4 + 166098*R*x**2 
- 155710*R - 180444*e*x**2 + 20000*e + 646247*x**4 
- 564311*x**2 + 105882.
\end{verbatim}

We take the same path as in the Hartree-Fock computation: the Gr\"obner basis, the primary ideal decomposition, and the numerical solution of the triangular system. The solutions are given in Table \ref{H3RX}.

\begin{table}[H]
\begin{center}
\begin{tabular}{rrrrrr}
\toprule
{} &          $x$ &            $e$ &        $R$ &  Total Energy   & Total Energy \\
{} &         {}  &            {} &        {} &  (Polynomial)   & (Exact) \\
\midrule
0  &  -0.5836 &      -14.2648 &  -1.4179 &          61.5478 &   \\
1  &   0.5836 &      -14.2648 &  -1.4179 &          61.5478 &   \\
2  &  -0.4053 &       -1.1465 &   1.8319 &          -1.2482 &  -1.2469 \\
3  &   0.4053 &       -1.1465 &   1.8319 &          -1.2482 & -1.2469  \\
4  &  -0.4510 &       -0.9208 &   2.5408 &          -1.1969 &    -1.1689  \\
5  &   0.4510 &       -0.9208 &   2.5408 &          -1.1969 &   -1.1689  \\
6  &  -0.8288 &        7.9352 &   4.6493 &         -24.5027 &   \\
7  &   0.8288 &        7.9352 &   4.6493 &         -24.5027 &   \\
8  &  -0.5227 &    -1584.5739 &  76.1157 &  -362449313.6859 &   \\
9  &   0.5227 &    -1584.5739 &  76.1157 &  -362449313.6859 &   \\
10 &  -0.0321 &  -321243.1293 &  76.1498 &  -278609641.4228 &   \\
11 &   0.0321 &  -321243.1293 &  76.1498 &  -278609641.4228 &   \\
\bottomrule
\end{tabular}
\caption
{
The real solutions of the polynomial equations for the optimized structure of H$_3^P$ are shown. Each row shows the values of x, e, R, and the corresponding total energy (the local optimum value of the polynomial objective function) in atomic units.
For comparison, the total energies evaluated by the exact analytic objective function are added to solutions No.2--5 which are the most plausible ones.
}
\label{H3RX}
\end{center}

\end{table}
Let us investigate the solutions. 

From the result, we have already removed the complex-valued solutions. We have only to check the real-valued ones. The first two solutions have negative $R$, and we judge that they are absurd and reject them. The solutions below No.6 have absurd total energies: they are located too deep. Now we get two types of solutions ( No.2,3 and No. 4, 5) that seem to be tolerable. 

Which of them could we trust? We know the optimum is unique. On account of the accuracy, we should choose the solutions with the values of R closer to the center of the Taylor expansion. According to Table \ref{H3RX},  the solutions of No.2 and 3 are located closer to the expansion center ($R_c=1.7\ a_0$) than No. 4 and 5. Hence we should choose the former.

There are measures to endorse the chosen solutions. First of all, we should check the accuracy of the polynomial approximation of the total energy. In a similar manner as in the previous section, using the exact analytic objective function, we can reiterate the total energy computations with various bond lengths. Table \ref{H3RX} includes the result of this check, which approves the solutions of No 2 and 3 as the optimum. Moreover, in the RHF computation by PySCF, the optimum is estimated at $R=1.8271\ a_0$ with the total energy -1.2469 $E_h$. It validates our computational strategy. 

Note that it is important to choose the center of the expansion properly. If we feel uncertain of the choice, we should compare the computations by setting different centers of the Taylor expansion so that the polynomial approximation would be sufficiently accurate.


\section{Notices on the decomposition of the ideal}
Here we give a brief account of a queer property of the solutions in Section \ref{SecRHF}: the orbital energies of the ground states and one of the excited states are computed simply from linear equations with integer coefficients. This is the consequence of our approximation, where we approximate every numeral coefficient by a rational number. The LCAO of the ground states is given by $v=(x,x,x)$, and that of the excited state under investigation is given by $v=(x,0,-x)$. They diagonalize the Fock and overlap matrix ($F$ and $S$) simultaneously, and the unknown x is determined by the normalization condition
\[
(x,x,x)\ S\left(\begin{array}{c} x\\x\\x \end{array}\right) =1
\] 
or
\[
(x,0,-x)\ S \left(\begin{array}{c} x\\0\\-x\end{array}\right) =1
\]
 
The orbital energies are given by 
\[
e=(x,x,x)\ F \left(\begin{array}{c} x\\x\\x \end{array}\right)
\] 
or
\[
e=(x,0,-x)\ F \left(\begin{array}{c} x\\0\\-x\end{array}\right)
\] 

Then we get the polynomial equation of x of the form $ S_2 x^2+S_0=1$ for the normalization and $E=F_4 x^4 + F_2 x^2+F_0$ for the orbital energies, for which all of the coefficients are rational numbers. If we approximate the entries of $S$ by rational numbers, $x^2$ is also a rational number. The computation of Gr\"obner basis does the elimination of $x^2$ and culls out the univariate polynomial of $E$ with integer coefficients that give the orbital energy as a rational number. 

\section{Notices on the optima of the total energy}
The optima of the total energy could be computed with constraints, even if the optima is not the solution of the Hartree-Fock equation in the strict sense. In Section \ref{SecRHF}, we allow the optimization of the total energy $E$ in the whole space $(x, y, z, e)$, and we did not get the wavefunction with the LCAO $(x,-2*x,x)$ as an optimum of the objective function. Nevertheless, we could get the minimum of the constrained case:
\[
E'(x,e)=\left.E(x,y,z,e) \right\lvert_{(x,y,z)=(x,-2x,x)}
\]
We show this minimum in Table \ref{optimax-2x-x}. Take note that this minimum is not the optimum of the total energy in the proper sense.
\begin{table}[H]
\begin{center}
\begin{tabular}{lrrr}
\toprule
{} &       x &       e & Total Energy \\
\midrule
0 & -0.6135 & -0.3988 &  0.2966 \\
1 & -0.6135 & -0.3988 &  0.2966 \\
\bottomrule
\end{tabular}
\end{center}
\caption{This table shows the optimum of the constrained objective function $E(x,-2x,x)$.}
\label{optimax-2x-x}
\end{table}

\section{
How to scrape out the excited states?
}
Using algebra, we scrape out the set of polynomial equations that determines the quantum state which we need. For example, we could extract excited states before we try the primary ideal decomposition; meanwhile, in Section \ref{SecRHF}, we get a ground state and two excited states through that treatment. The required trick is based on the {\it ideal quotient}. 
Let $R$ be the base ring; $I$, $J$ be ideals; and $M$ a module in ${\tt R}^n$. Then the ideal quotient is defined by
\[
quotient(I,J)= \{a \in R \mid aJ \subset I\},
\]
Recall the zeros of the ideal I and J depict the corresponding geometric objects V(I) and V(J) in the Cartesian coordinate. Then The quotient(I, J) cuts the figure $V(I)\setminus V(J)$, namely, the remaining part of V(I) after the removal of $V(I)\cap V(J)$.

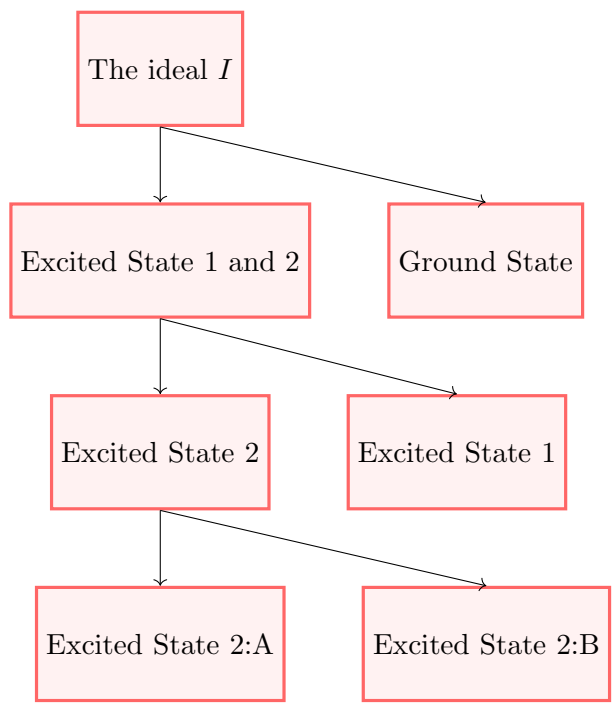
\begin{figure}[H]
\begin{center}
    
\begin{tikzpicture}[
roundnode/.style={circle, draw=green!60, fill=green!5, very thick, minimum size=17mm},
squarednode/.style={rectangle, draw=red!60, fill=red!5, very thick, minimum size=15mm},
]
\node[squarednode]      (excite12) {
Excited State 1 and 2
};
\node[squarednode]        (ideal)       [above=of excite12] {
The ideal $I$
};
\node[squarednode]      (ground)       [right=of excite12] {
Ground State
};
\node[squarednode]        (excite2)       [below=of excite12] {
Excited State 2
};
\node[squarednode]        (excite1)       [right=of excite2] {
Excited State 1
};
\node[squarednode]        (excite2A)       [below=of excite2] {
Excited State 2:A
};
\node[squarednode]        (excite2B)       [right=of excite2A] {Excited State 2:B
};

\draw[->] (ideal.south) -- (ground.north);
\draw[->] (ideal.south) -- (excite12.north);

\draw[->] (excite12.south) -- (excite2.north);
\draw[->] (excite12.south) -- (excite1.north);
\draw[->] (excite2.south) -- (excite2A.north);
\draw[->] (excite2.south) -- (excite2B.north);


\end{tikzpicture}
\end{center}

\caption{
The triangular decomposition of the ideal I into the subsets, each of which depicts various quantum states. The ideal I is decomposed into two subsets (the ground state and the union of excited states 1 and 2). The union of two excited states is then separated into distinct states. On account of the algorithm of symbolic computation, the second excited state is furthermore decomposed into two triangular sets (2:A and 2:B).
}
\label{TriangleDecomp}
\end{figure}

Let $I$ be the ideal that defines the quantum states of H$_3^+$. We know the following items:
\begin{itemize}
\item The ground state is the intersection of $I$ and another ideal $J1=(x-y, y-z)$.
\item A type of excited state is featured by the zero at one of x, y, or z. Hence this type of excited state is the intersection of $I$ and an ideal $J2=(xyz)$.
\item The decomposition of the ideal I into distinct quantum states is depicted in Figure \ref{TriangleDecomp}. 
\end{itemize}

For this reason, we get the two excited states as the zeros of the quotient ideal: Q1=Q(I, J1). Moreover, we get one excited state by Q2=Q(Q1, J2), by removing the component of another excited state. In the following, we show Q1 and Q2. The results affirm the ability of the ideal quotient.

In each level of the decomposition of the ideals in Figure \ref{TriangleDecomp}, the polynomial which gives the orbital energies goes through the following change:   
\begin{verbatim}
Ideal I / Ground and excited states:
36812798143709139749144994785898166628646912*e^5
+127499329072859314688739373550168339532546048*e^4
+168144765973871153972385957671411273274218496*e^3
+105507004438312720335072205511167338086667520*e^2
+31442209826685795119941751238920406402581000*e
+3570217123261998609632303137992216382496875

Excited States 1 and 2:
658740729187721858486566052982030336*e^4
+1491545236264568515424785662354382848*e^3
+1220157666177907747137353882531720960*e^2
+424748038399486631113783014582688000*e
+53273792919663046138734741668365625

Excited States 2:
33596906582185481529961979904*e^3
+62671838378960619969574479872*e^2
+37234589977346832313644870640*e
+6812527267627204196782565375
\end{verbatim}

In Figure \ref{TriangleDecomp}, we have two subsets (2:A and 2:B) for the excited state 2. They have a common polynomial which determines the orbital energies and is shown in the above. The difference between the subsets 2:A and 2:B lies in the polynomials located underneath (at the second and the following ones) in their triangular set. (Cf. the primary ideal decomposition which we computed in Section \ref{SecRHF}.)

Note that if there is no intersection between two ideals $I$ and $J$, then $I=quotient(I, J)$, namely, $I$ is kept intact. We again assume that $I$ is the ideal that defines the quantum states of H$_3^+$. We choose the ideal $J=(x-y,z+2x$) such that the corresponding LCAO is $(x,x,-2x)$. The computational experiment shows that $quotient(I, J)$ returns the Gr\"obner basis of $I$. It means that the quantum states of the LCAO $(x,x,-2x)$ are not the eigenstate of the Hamiltonian described by $I$, meanwhile, the wavefunctions with the LCAO $(x,x,x)$, $(x,x,0)$, $(x,0,x)$ and $(x,x,0)$ diagonalize the Hamiltonian.

\section{
The computation of the virtual orbitals in the Hartree-Fock model
}
We can compute the virtual orbitals orthogonal to the Hartree-Fock molecular orbital $(x,y,z)$. The virtual orbitals are the eigenvectors of the Fock matrix $F_{GR}$ whose non-linear entries are determined by the ground state. The computation demands us a bit of care. 

The objective function to determine the virtual orbitals by $(u,v,w)$ is given by
\begin{align*}
&(u,v,w) F_{GR} \left(\begin{array}{c} u\\v\\w \end{array}\right)
-s\left(
(u,v,w) S \left(\begin{array}{c} u\\v\\w \end{array}\right) -1
\right)\\
&-t\left(
(u,v,w) S \left(\begin{array}{c} x\\y\\z \end{array}\right) 
\right)\\
&+\left( H_{tot}(x,y,z)-e\left( (x,y,z)\ S \left(\begin{array}{c} x\\y\\z \end{array}\right)-1\right)  \right),
\end{align*}
where $H_{tot}$ is the objective function for the Hartree-Fock wavefunction (x,y,z), and the constraints for the normalization and the orthogonalization are provided.

To get the solutions, it is wise to apply divide-and-conquer: only if the efficiency of the hardware is sufficiently high, we could follow the similar steps demonstrated in the previous section. We replace the undermined $(x,y,z)$ with a fixed numeral vector $(x_v, x_v, x_v)$ that represents the Hartree-Fock ground state. In this case, $H_{tot}$ is also a constant, and the last term in the objective function is neglected. Then the approximated objective function necessary to determine the virtual orbital is given by
\begin{verbatim}
OBJ=
-10000*s*u**2 -11144*s*u*v-11144*s*u*w-10000*s*v**2
-11144*s*v*w-10000*s*w**2+10000*s+8395t*u+8395t*v
+8395t*w-8548*u**2-16808*u*v-16808*u*w-8548*v**2
-16808**vw-8548*w**2.
\end{verbatim}
To get it, we substitute the numerical result of the ground state to $x_v$ and $e$, approximate the digit coefficients by rational integers and multiply the objective function by the common denominator.  Computing partial derivatives,  we get the ideal which gives the minima. It is composed of the following polynomials:
\begin{verbatim}
[
-10000*u2-11144*u*v-10000*v**2-11144*u*w-11144*v*w-10000*w**2+10000,
8395*u+8395*v+8395*w,
-20000*s*u-11144s*v-11144*s*w+8395*t-17096*u-16808*v-16808*w,
-11144*s*u-20000s*v-11144*s*w+8395*t-16808*u-17096*v-16808*w,
-11144*s*u-11144s*v-20000*s*w+8395*t-16808*u-16808*v-17096*w
]
\end{verbatim}

The Gr\"obner basis for this ideal is computed as follows.
\begin{verbatim}
{
 u+v+w,
 t,
 123*s+4,
 1107*v**2+1107*v*w+1107*w**2-1250
}.
\end{verbatim}

Unfortunately, the dimension of the ideal is 1; this means the degeneracy of the virtual orbitals. In other words, the roots of the corresponding polynomial set are not isolated; we merely get a one-dimensional geometrical object (an algebraic curve) through polynomial relations between the variables. Nevertheless, we get the unique eigenvalue at $s$ determined by a linear univariate polynomial $123s+4$, and this eigenvalue is common to all solutions given on this algebraic curve. 

Note that, if we use an additional condition, say, $u=0$, we get the zero-dimensional ideal, and we get the isolated solution of a virtual orbital. It seems that the ideal quotient might help us to get another virtual orbital.  In the current case, the minimum of the objective function is an algebraic curve defined by the ideal I, and the additional condition $u=0$ is given by another ideal J.  Then $I\cap J$ is a zero-dimensional ideal, and the computation of the $quotient(I, J)$ is possible. However, it does not serve the aim to get another virtual orbital. $quotient(I, J)$ is the closure of $V(I)\setminus V(I\cap J)$, hence it still has dimension one and the corresponding algebraic curve is intact. That is to say, we have got stuck at $I=quotient(I, J)$.

It would be too naive to substitute a digit to the ground state wavefunction $(x,x,x)$, and we could use a more complicated objective function. Recall that the ground-state wavefunction is determined by 
$
(x,x,x)S\left(\begin{array}{c} x\\y\\z \end{array}\right)=1.
$
Then we construct the objective function for the virtual orbitals $(u,v,w)$ with constraints as follows: 
\begin{align*}
&(u,v,w) F_{GR} \left(\begin{array}{c} u\\v\\w \end{array}\right)
-e_v\left(
(u,v,w) S \left(\begin{array}{c} u\\v\\w \end{array}\right) -1
\right)\\
&+r\left(
(u,v,w) S \left(\begin{array}{c} x\\y\\z \end{array}\right) 
\right)
+q\left( (x,y,z)\ S \left(\begin{array}{c} x\\y\\z \end{array}\right)-1\right),
\end{align*}
It determines both the occupied and virtual orbitals in the ground state. The polynomial approximation is given by

\begin{verbatim}
OBJ=
63431*q*x**2-10000*q-10000*s*u**2-11144*s*u*v
-11144*s*u*w-10000*s*v**2-11144*s*v*w
-10000*s*w**2+10000*s-21144t*u*x-21144t*v*x
-21144t*w*x+46573*u**2*x**2-15891*u**2
+34956*u*v*x**2-22319*u*v+34956*u*w*x**2
-22319*u*w+46573*v**2*x**2-15891*v**2
+34956*v*w*x**2-22319*v*w+46573*w**2*x**2
-15891*w**2
\end{verbatim}

The optimum of this objective function is also computed from an ideal that is composed of the partial derivatives of the objective function.  This ideal is of dimension one and its Gr\"obner basis for this ideal is computed as follows:
\begin{verbatim}
u+v+w,
t,
20805368*s+679539,
70218117*q+72737500,
1107*v**2+1107*v*w+1107*w**2-1250,
63431*x**2-10000.
\end{verbatim}

Note that there is a slight difference in the polynomial ideals in the former and the latter cases. This is attributed to the numerical error that happened in the substitution. In the former case, we substitute a digit number to the $x_v$ in the formula and then approximated the coefficients by rational numbers. Meanwhile, in the latter case, we gave the $x_v$ through a polynomial. The latter approach would be preferable since it shall suppress the occurrence of extra numerical errors. 

\section{
Hartree-Fock calculation by applying eigensolver only once
}

We could conduct the Hartree-Fock computation by applying the eigenvalue solver only once, without entering iterations to achieve self-consistency.  This type of computation is available from the following ground.

Let us consider the factor ring $Q=R[x,y,z,e]/I$ with the ideal $I$ that determines the electronic structure of H$_3^+$. As $I$ is zero-dimensional, $Q$ is generated by a finite set of the power products of the variables. We call those generators q-base. In the current case, the q-base is composed of 26 entries as follows:
\begin{verbatim}
b[1]=y*z^2 b[2]=y*z*e^3 b[3]=y*z*e^2 b[4]=y*z*e
b[5]=y*z b[6]=y*e^3 b[7]=y*e^2 b[8]=y*e b[9]=y
b[10]=z^3*e^2 b[11]=z^3*e b[12]=z^3 b[13]=z^2*e^3
b[14]=z^2*e^2 b[15]=z^2*e b[16]=z^2 b[17]=z*e^4
b[18]=z*e^3 b[19]=z*e^2 b[20]=z*e b[21]=z 
b[22]=e^4 b[23]=e^3 b[24]=e^2 b[25]=e
b[26]=1
\end{verbatim}

In the quotient ring, any multiplication of an element over the q-base is a linear transformation over there.  We can compute the linear transformation matrices that represent the multiplication of $x, y, z$, and $e$ over the q-base. The eigenvalues of those matrices are the roots of the set of polynomial equations defined by the ideal $I$. As we work in commutative algebra, those matrices are commutative among themselves and share the common eigenvectors $\{v_i|i=1,.., N\}$. Let $M(x), M(y), M(z)$, and $M(e)$ be the transformation matrices that represent the multiplication to the q-base by $x, y, z,$ and $e$.
The eigenvalues are determined by the following relations,
\begin{align*}
M(x)\ v_i= x_i\ v_i \\
M(y)\ v_i= y_i\ v_i \\
M(z)\ v_i= v_i\ v_i \\
M(e)\ v_i= e_i\ v_i \\
\end{align*}
We get the set of the roots $(x_i,y_i,z_i,e_i)$ from each of $v_i (i=1,...,N)$.  There is a tricky part in obtaining the eigenvectors which diagonalize all of the linear transformation matrices simultaneously. If there is a degeneracy of the eigenvalues of a linear transformation matrix,  the eigenvectors for an eigenvalue are not unique, and a representative of the eigenspace might not be the proper eigenvector of other linear transformation matrices. Moreover, the entries in $v_i (i=1,...,N)$ represent the q-base after the substitution by $(x,y,z,e)=(x_i,y_i,z_i,e_i)$. Hence, in the choice of the eigenvector, the entry on the slot, where b[...]=1 is located, should not be zero. 

To evince the power of this method, let us try to calculate the eigenvalue of $M(e)$. Table \ref{matrixMe} shows the real eigenvalues and they agree with the result by triangulation.
 
\begin{table}[H]
\begin{center}
    
\begin{tabular}{lll}
\toprule
{} &                                                Eigenvalue & Multiplicity \\
\midrule
 &                                            -0.3531 &   6 \\
 &                                            -0.3988 &   6 \\
 &                                            -1.1992 &   2 \\
\bottomrule
\end{tabular}
\end{center}

\caption{
The real eigenvalues of the linear transformation matrix that represents the action of the variable $e$ over the q-base are shown with their multiplicities. The numerical computation is executed by the eigensolver implemented in SINGULAR.
} 
\label{matrixMe}
\end{table}

\section{
Discussion: a possible linkage to quantum computation
}
We discuss the contact between the algebraic method of quantum chemistry presented in this article and the quantum computation that is in vogue now. 

\begin{figure}[H]
\begin{center}

\begin{quantikz}
\lstick{$\ket{0}$}& \gate{H} & \ctrl{4} & \qw&\ \ldots\ \qw & \qw & \gate[4]{QFT^\dagger} &
\qw &\meter{}\\
\lstick{$\ket{0}$}& \gate{H} & \qw & \ctrl{3}&\ \ldots\ \qw & \qw & \qw &
\qw &\meter{}\\
\wave&&&&&&&&\\
\lstick{$\ket{0}$}& \gate{H} & \qw & \qw &\ \ldots\ \qw & \ctrl{1} & \qw &
\qw &\meter{}\\
\lstick{$\ket{g}$}& \qwbundle{\ } & \gate{U^{2^0}} & \gate{U^{2^1}} &\ \ldots\ \qw & \gate{U^{2^k}} & \qw & \qw&\lstick{$\ket{g}$}
\end{quantikz}
\end{center}
\caption{
The quantum circuit for quantum phase estimation.
The top registers ($|0\rangle^{\otimes t}$) are used for counting, being followed by Hadamard gates $H$, whereas the bottom resister holds the state $|g\rangle$. Through controlled unitary operations $\{U^{2^k},k=0,...,t\}$ and the inverse quantum Fourier transformation $QFT\dagger$, the eigenvalue of the unitary operator $U$ on $g\rangle$ is computed as the integer representation of n-bit binary numbers, which are component-wise obtained by the measurements at the right end of the circuit.
}
\label{QPEFIG}
\end{figure}
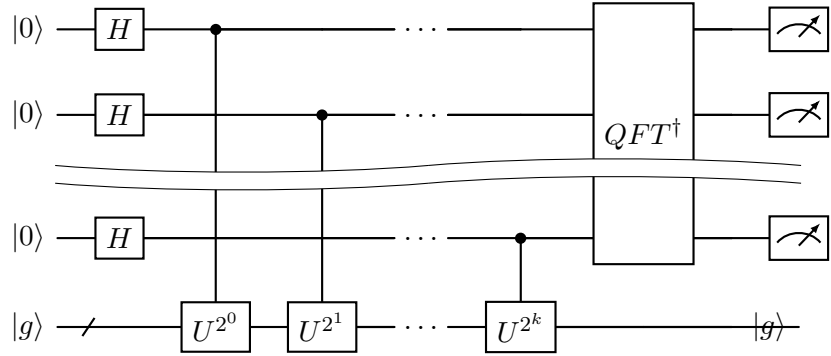

The plausible connection between them is evolved from the following points.
\begin{itemize}
\item 
One of the featured algorithms in quantum computation is quantum phase estimation (QPE), which enables us to evaluate the eigenvalue of a Hermitian operator (i.e. Hamiltonian) through time evolution. The eigenvalue of the target Hamiltonian is obtained at the phase factor of the quantum state processed by the quantum circuit depicted in Figure \ref{QPEFIG}.
\item 
The algebraic method demonstrated in the present article enables us to solve the Hartree-Fock computation through the eigenvalue problem of the transformation matrices of the q-bases.
\item
Hence, it seems that QPE shall be useful to solve the eigenvalue problem in the q-bases. 
\end{itemize}

If we express this anticipation through the language of quantum mechanics, it goes as follows.
Let $\{M_i, i=1,..., N\}$ be the transformation matrices that represent the action of the indeterminate variables $\{x_i,i=1,..., N\}$ on the q-bases, and let $\{w^l, l=1,..., M\}$ be the common eigenvectors to 
$\{M_i, i=1,...,N\}$. We might conduct a quantum computation in the following way.

\[
|w^l\rangle \rightarrow |w^l\rangle |(w^l|M_1|w^l)\rangle \cdots |(w^l|M_N|w^l)\rangle
\] 

Therefore we could write the eigenvalues of $\{M_i, i=1,..., N\}$ on the ancilla q-bits.

In pity, it is impossible to apply quantum computation to the present algebraic method directly. This is on account of the following reasons.
\begin{itemize}

\item The quantum phase estimation is applied to the unitary operator that is the time-evolution by the hermitian Hamiltonian: $\exp(i\cdot t\cdot H)$.
\item The transformation matrices M in the quotient ring are not symmetric, hence not hermitian. They should not be regarded as the usual operators in quantum mechanics. There can be no normal quantum models that are propagated by those matrices.
\end{itemize}

To get over this discordance, we would employ the following tricks.

Let $A$ and $V$ be the operators, defined by complex-valued matrices. Additionally, let us assume that $A$ is included in $V$ as a submatrix at the diagonal part in such a way that
\[
V=\begin{pmatrix}
A & * \\
* & *
\end{pmatrix}
\]

With the aid of projection, we could compute the phase for the relevant part. Then $A$ is block encoded into $V$ by the signal states $\{ |\mathcal{L}\rangle\}$, provided that $V$ is chosen to be a unitary operator. One would put at $A$ an arbitrary bounded-norm operator. With the projection operator $\Pi=\sum_{\mathcal{L}}|\mathcal{L}\rangle\langle\mathcal{L}|$, the block encoding is written by $\Pi V \Pi = A$. This representation enables us to operate an arbitrary $A$ on qubits if we embed $A$ into $V$ and operate $V$ on a larger system extended by ancillary qubits. There are various methods to apply quantum phase estimations on arbitrary operators with the aid of block encoding \cite{yoshioka2022hunting}, such as qubitization \cite{low2019hamiltonian} or Taylorization\cite{gilyen2019quantum}. This sort of algorithm shall serve as the preparation of initial quantum states in quantum computations for Post-Hartree-Fock methods. This treatment shall increase the computational cost over the innate computational complexity regarding the algebraic geometry. Nevertheless, the quantum machine could write the information of the quantum state on its tape, provided that the eigenvalues (namely, the corresponding phase shifts) are computed for all of the required transformation matrices in the quotient ring.

Note that it is probable that the quantum computation of eigenvalues of the transformation matrix in the quotient ring would involve other technical difficulties: for example, the transformation matrices would include complex-valued eigenvalues, which shall make the time evolution unstable. We might anticipate the damping of the quantum states with complex-valued energy, but this is not the expected physics in quantum circuits. It seems that we should make a compromise. Therein we compute the real eigenvalues and the corresponding eigenvectors for one of the transformation matrices by the classical methods, and then we employ the quantum algorithm to evaluate the eigenvalues for other transformation matrices. (Recall that those transformation matrices share the eigenvectors.)

\section{
Summary and conclusion
}
In this article, we demonstrated the restricted Hartree-Fock computation of the electronic structure of H$_3^+$. We use the approximation of analytic energy functional by a polynomial, and then get a set of polynomial equations to determine the optima. We modify and rearrange the polynomials through the technique of computational algebraic geometry so that we shall get the solutions with ease. We have seen the power of the Gr\"obner basis, the triangulation, the ideal quotient, and related tricks, which enable us to study the electronic structure of polyatomic molecules. The complexity of the computations in the present article is not so large that we could execute them on small desktop computers. However, the complexity of the unrestricted-Hartree-Fock (UHF) computation of H$_3^+$ grows enormous. The authors of the current article could not complete this computation. Hence it would be an open problem for the readers, and the computer program of the UHF case is available at GitHub \cite{OURGITHUB}.  An important future theme for us is to take into consideration the many-body interactions beyond the Hartree-Fock model. It shall also be the target of algebraic studies, so long as we follow the standard formalism of quantum chemistry, wherein problems are represented by ubiquitous polynomials. It is possible to construct the polynomial representation of the total energy for unrestricted-Hartree-Fock computation of H$_3^+$. 

\bibliographystyle {unsrt}
\bibliography{biball}
\input{appendix}

\section*{Supplementary materials}
We made public the raw results of the computations. One can get them from the following place:
\begin{verbatim}
http://github.com/kikuchiichio/AlgebraicQuantumChemistry
\end{verbatim}
\end{document}

%% file: appendix.tex
\newpage
\section*{Appendix: A proof of the existence of solutions of RHF model of H$_3^+$.}

In this appendix, we give brute-force proof of the existence (and the non-existence) of the solutions of the RHF model of H$_3^+$. In the computations presented in the main part of the article, we used concrete numeral molecular integrals. We have seen that several types of solutions ($(x,x,x),(0,x,x), (x,0,x), (x,x,0)$) exist, while the others ($(-2x,x,x),(x,-2x,x),(x,x,-2x)$) not. We shall verify this is generally the case, irrespective of the choice of the numerical data.

Let us assume that H$_3^+$ is made of three s-orbitals $(i=0,1,2)$ on the vertices of an equilateral triangle.

For simplicity, we write the one-body part of the Fock and overlap matrices as follows:
\[
\left(\begin{array}{ccc} 0 & H & H \\ H & 0 & H \\ H & H & 0 \end{array}\right), 
\left(\begin{array}{ccc} 1 & S & S \\ S & 1 & S \\ S & S & 1 \end{array}\right), 
\]
We remove the diagonal part of the one-body Fock matrix because it merely causes the shift of the molecular orbital energy.

We write the wave function and the orbital energy by $(x,y,z)$ and $e$. 

The two-body part of the Fock matrix is composed of the two-electron integrals between four centers: [I, J|K, L]. Recall that there is an equivalence of those integrals, due to the definition, such that [I, J|K, L]=[J, I|K, L]=[I, J|L, K]=[K, L|I, J] and so on. In addition, there is another equivalence owing to the triangular symmetry. For this reason, to construct the energy functional, we have only to use the following six two-electron integrals:
\begin{verbatim}
(TT0000,TT0001, TT011, TT0012, TT0101, TT0102)
= ([00|00], [00|01] ,[00|11], [00|12], [01|01], [01|02])
\end{verbatim}

The objective function is given by 
\begin{verbatim}
f=4*H*x*y + 4*H*x*z + 4*H*y*z - 4*S*e*x*y - 4*S*e*x*z
- 4*S*e*y*z + TT0000*x**4 + TT0000*y**4 + TT0000*z**4
+ 4*TT0001*x**3*y + 4*TT0001*x**3*z + 4*TT0001*x*y**3
+ 4*TT0001*x*z**3 + 4*TT0001*y**3*z + 4*TT0001*y*z**
3 + 2*TT0011*x**2*y**2 + 2*TT0011*x**2*z**2 
+ 2*TT0011*y**2*z**2 + 4*TT0012*x**2*y*z 
+ 4*TT0012*x*y**2*z + 4*TT0012*x*y*z**2 
+ 4*TT0101*x**2*y**2 + 4*TT0101*x**2*z**2 
+ 4*TT0101*y**2*z**2 + 8*TT0102*x**2*y*z 
+ 8*TT0102*x*y**2*z + 8*TT0102*x*y*z**2 
- 2*e*x**2 - 2*e*y**2 - 2*e*z**2 + 2*e
\end{verbatim}
The optima of the objective function are determined by the ideal  $$I=(df/dx, df/dy, df/dz, df/dx),$$ which is defined as follows.
\begin{verbatim}
I[1]=4*x^3*TT0000+12*x^2*y*TT0001+4*y^3*TT0001
+12*x^2*z*TT0001+4*z^3*TT0001+4*x*y^2*TT0011
+4*x*z^2*TT0011+8*x*y*z*TT0012+4*y^2*z*TT0012
+4*y*z^2*TT0012+8*x*y^2*TT0101+8*x*z^2*TT0101
+16*x*y*z*TT0102+8*y^2*z*TT0102+8*y*z^2*TT0102
-4*y*e*S-4*z*e*S-4*x*e+4*y*H+4*z*H

I[2]=4*y^3*TT0000+4*x^3*TT0001+12*x*y^2*TT0001
+12*y^2*z*TT0001+4*z^3*TT0001+4*x^2*y*TT0011
+4*y*z^2*TT0011+4*x^2*z*TT0012+8*x*y*z*TT0012
+4*x*z^2*TT0012+8*x^2*y*TT0101+8*y*z^2*TT0101
+8*x^2*z*TT0102+16*x*y*z*TT0102+8*x*z^2*TT0102
-4*x*e*S-4*z*e*S-4*y*e+4*x*H+4*z*H

I[3]=4*z^3*TT0000+4*x^3*TT0001+4*y^3*TT0001
+12*x*z^2*TT0001+12*y*z^2*TT0001+4*x^2*z*TT0011
+4*y^2*z*TT0011+4*x^2*y*TT0012+4*x*y^2*TT0012
+8*x*y*z*TT0012+8*x^2*z*TT0101+8*y^2*z*TT0101
+8*x^2*y*TT0102+8*x*y^2*TT0102+16*x*y*z*TT0102
-4*x*e*S-4*y*e*S-4*z*e+4*x*H+4*y*H

I[4]=-4*x*y*S-4*x*z*S-4*y*z*S-2*x^2-2*y^2-2*z^2+2
\end{verbatim}
The Gr\"obner basis of the ideal I is without difficulty computed in the ring $\mathbb{Q}$[x, y, z, e, H, S, TT0000, TT0001, TT011, TT0012, TT0101, TT0102]. The ring that is the most convenient for the computation is given by twelve variables with the degree reverse lexicographic monomial ordering:
\begin{verbatim}
x > y > z > e > H > S > TT0000 > TT0001 > TT011 > TT0012 > TT0101 > TT0102
\end{verbatim}
That Gr\"obner basis is composed of 56 polynomials and its dimension is eight.  It means that if we arbitrarily choose eight variables in the rings and substitute numeral values for them, we get the optima as isolated solutions. That is to say, if we give the concrete values for eight molecular integrals (H, S, TT0000, TT0001, TT011, TT0012, TT0101, TT0102), the possible electronic structures are uniquely determined.

Let us give some constraints to the model. 

To get the solution $(x,x,x)$, we extend the ideal $I$ by two polynomials $y-x$ and $z-x$ and get:
\begin{verbatim}
 J1={I[1],I[2],I[3],I[4], x-y, z-x}.
\end{verbatim}
The Gr\"obner basis of J1 is easily computed and we get
\begin{verbatim}
_[1]=y-z
_[2]=x-z
_[3]=z^2*TT0000+8*z^2*TT0001+2*z^2*TT0011+4*z^2*TT0012
     +4*z^2*TT0101+8*z^2*TT0102-2*e*S-e+2*H
_[4]=12*e*S^2+12*e*S-12*H*S+3*e-6*H-TT0000-8*TT0001
     -2*TT0011-4*TT0012-4*TT0101-8*TT0102
_[5]=6*z^2*S+3*z^2-1
\end{verbatim}
This ideal is of dimension 8. If we substitute the concrete numeral values for eight molecular integrals, the wavefunction $(x,x,x)$ is determined.

To get the solution $(x,0,-x)$, we extend the ideal $I$ by adding two polynomials $y$ and $z+x$ and get:
\begin{verbatim}
J2={I[1],I[2],I[3],I[4], y, z+x}.
\end{verbatim}
The Gr\"obner basis of J2 is easily computed and we get
\begin{verbatim}
_[1]=y
_[2]=x+z
_[3]=z^2*TT0000-4*z^2*TT0001+z^2*TT0011+2*z^2*TT0101+e*S-e-H
_[4]=2*e*S^2-4*e*S-2*H*S+2*e+2*H-TT0000+4*TT0001-TT0011-2*TT0101
_[5]=2*z^2*S-2*z^2+1
\end{verbatim}
This ideal is of dimension 8. If we substitute the concrete numeral values to eight molecular integrals the wavefunction $(x,0, -x)$ is determined.

As for the solution $(x, -2x,x)$, we prepare the ideal:
\begin{verbatim}
J3={I[1],I[2],I[3],I[4], y+2x, z-x}.
\end{verbatim}
The Gr\"obner basis of J3 is composed of the following polynomials
\begin{verbatim}
_[1]=TT0000-TT0001-TT0011+TT0012-2*TT0101+2*TT0102
_[2]=y+2*z
_[3]=x-z
_[4]=9*z^2*TT0001-6*z^2*TT0011+3*z^2*TT0012-12*z^2*TT0101+6*z^2*TT0102-e*S+e+H
_[5]=2*e*S^2-4*e*S-2*H*S+2*e+2*H+3*TT0001-2*TT0011+TT0012-4*TT0101+2*TT0102
_[6]=6*z^2*S-6*z^2+1
\end{verbatim}
The ideal J3 is of dimension 7, and we could not freely provide the equations with eight molecular integrals. The first polynomial of the Gr\"onber basis is made only of the two-electron integrals TT0000, TT0001, TT0011, TT0012, TT0101, and TT0102. If the problem under investigation has any solution, those six two-electron integrals should make this polynomial zero. However, this is impossible for general cases. That is to say, the wavefunction $(x,-2x,x)$ is not allowed to be the solution of the RHF model of H$_3^+$.

To get the solution $(x,y,x)$ such that $y\ne x$, we compute the ideal quotient. 
Let
\begin{verbatim}
J4={I[1],I[2],I[3],I[4], z-x},
K= {y-x},
\end{verbatim}
and
\begin{verbatim}
J5=quotient(J4,K)
\end{verbatim}
The Gr\"obner basis of J5 is of dimension 8, containing 16 polynomials.  We omit its complicated expression. This ideal gives the proper eigensolution if we give eight molecular integrals.